\documentclass[article, sort&compress]{elsarticle}

\usepackage{hyperref}
\usepackage{graphicx}
\usepackage{amsmath}
\usepackage{amssymb}
\usepackage{color}
\usepackage[normalem]{ulem}
\usepackage{lineno}

\def\OA{{(\Omega, A)}}

\begin{document}

\begin{frontmatter}

\title{Supratransmission-induced travelling breathers in long Josephson junctions}


\author[unipa]{Duilio De Santis\corref{correspondingauthor}}
\cortext[correspondingauthor]{Corresponding author}
\ead{duilio.desantis@unipa.it}
\author[unisa]{Claudio Guarcello}
\author[unipa,unn]{Bernardo Spagnolo}
\author[unipa]{Angelo Carollo}
\author[unipa]{Davide Valenti}
\address[unipa]{Dipartimento di Fisica e Chimica ``E.~Segr\`{e}", Group of Interdisciplinary Theoretical Physics, Università degli Studi di 
Palermo, I-90128 Palermo, Italy}
\address[unisa]{Dipartimento di Fisica ``E.~R.~Caianiello", Università degli Studi di Salerno, I-84084 Fisciano, Salerno, Italy}
\address[unn]{Radiophysics Department, Lobachevsky State University, 603950 Nizhniy Novgorod, Russia}

\begin{abstract}
The emergence of travelling sine-Gordon breathers due to the nonlinear supratransmission effect is theoretically studied in a long Josephson junction driven by suitable magnetic pulses, taking into account the presence of dissipation, a current bias, and a thermal noise source. The simulations clearly indicate that, depending on the pulse's shape and the values of the main system parameters, such a configuration can effectively yield breather excitations only. Furthermore, a nonmonotonic behavior of the breather-only generation probability is observed as a function of the noise intensity. Finally, the dynamics of the supratransmission-induced breathers is characterized by looking at quantities such as their radiative decay lifetime and the medium's energy.
\end{abstract}


\end{frontmatter}

\section{Introduction}
\label{Sec1}

The sine-Gordon~(SG) equation is commonly assumed to provide a good description of the properties of a long Josephson junction~(LJJ)~\cite{Barone_1982}. This nonlinear hyperbolic partial differential equation appears in a wide range of research fields~\cite{Ivancevic_2013, Jesus_2014, Bykov_2014, Villari_2018, Yakushevich_2021}, and it is typically regarded as one of the three fundamental soliton equations, the others being the Korteweg-de Vries equation and the nonlinear Schr\"{o}dinger equation~\cite{Scott_2003, Dauxois_2006}. As a result, the LJJ has become a prototypal solid-state device for the exploration of soliton dynamics~\cite{Parmentier_1993, Ustinov_1998}.

More precisely, perturbative correction terms to the pure SG equation arise when the effects of dissipation, an external current source, and thermal fluctuations are included to properly model the response of a realistic junction. Introduced as exact solutions of the original equation, the soliton profiles describe robust, particle-like objects which tend to maintain their identity, to some extent, in the perturbed case as well. Hence, both analytical and numerical investigations concerning the dynamics of these nonlinear excitations under different circumstances have attracted significant research efforts over the years~\cite{McLaughlin_1978_PRA, Kivshar_1989, Lomdahl_1986, Jensen_1992, Gulevich_2012, Gulevich_2006, Johnson_2013, De_Santis_2022}.

The simplest SG soliton, i.e., the kink, has been the subject of extensive studies also in the presence of deterministic and stochastic perturbations~\cite{McLaughlin_1978_PRA, Kivshar_1989}. Such a solitary wave, whose stability is truly remarkable, has a clear physical interpretation in the context of LJJs, since it exactly carries a quantum of magnetic flux~$ \Phi_0 $, induced by a supercurrent loop surrounding it (for this reason, the kink is often referred to as fluxon or vortex). Kinks are promptly trackable as voltage steps on the \textit{I}-\textit{V} characteristics of the junction, and since they can be stored, steered, manipulated, and made to interact with electronic instrumentation, their features are exploited in many applications~\cite{Parmentier_1993, Ustinov_1998, Soloviev_2014, Soloviev_2015, Guarcello_2017, Gua18, GuaSolBra18, Wustmann_2020, Osborn_2021}. Nonetheless, the interplay between the system's temperature and the evolution of the mode has been demonstrated. In particular, thermal fluctuations can influence the vortex's behavior~\cite{Castellano_1996, Fedorov_2007, Fedorov_2008, Aug09}, and even a temperature bias can affect both the soliton dynamics~\cite{Log94,Gol95,Kra97} and the thermal transport across the device~\cite{Gua18,GuaSolBra18,GuaSol18,Gua19}.

The attractive interaction between a kink and an antikink (a sign distinguishes the latter from the kink solution) can lead to the formation of a bound state, known as \emph{breather} (or bion, in the earlier literature), which is space-localized and time-periodic. Breathers are far more elusive objects than kinks, mainly because they radiatively decay due to dissipation~\cite{McLaughlin_1978_PRA}, unless specific forms of driving are employed~\cite{Lomdahl_1986, Jensen_1992}, and because their oscillatory nature gives rise to a practically null average voltage, i.e., beyond the sensitivity of the existing high-frequency oscilloscopes~\cite{Gulevich_2012}. As a consequence, despite the large variety of studies devoted to them~\cite{McLaughlin_1978_PRA, Kivshar_1989, Lomdahl_1986, Jensen_1992, Gulevich_2012, Gulevich_2006, Johnson_2013, De_Santis_2022}, definitive experimental evidence has yet to be found in LJJs. Indeed, the observation of breather modes in Josephson systems has only been reported for rotobreathers in Josephson ladders~\cite{Trias_2000, Binder_2000, Segall_2014}.

The detection of breathers in LJJs constitutes an intriguing open challenge in the realm of mesoscopic soliton physics, but the interest towards such an excitation is also motivated by its significant applicative potential. In fact, this nonlinear wave could be effectively used to develop some novel applications in information transmission~\cite{Macias-Diaz_2007, Macias-Diaz_2007_1}. Also, in contrast to kinks and antikinks, breather modes possess an internal degree of freedom, i.e., a proper frequency, which is particularly valuable for quantum computation purposes. More specifically, breathers behave as macroscopic artificial two-level atoms in an LJJ with a small capacitance per unit length, so that the realization of a Josephson breather qubit has been proposed~\cite{Fujii_2007, Fujii_2008, Fuj09}. Such a massive mobile qubit would act as a quantum data bus transferring quantum information between different nodes, while being able to perform calculations during communication without the support of stationary qubits. Thus, since multiple aspects concerning efficient breather generation and control setups remain to be clarified, further advancements are also relevant in view of concrete applications.

It should be mentioned that topologically neutral nonlinear waves, such as fluxon-antifluxon couples and breathers, can play a significant role in a Kibble-Zurek-like scenario~\cite{Monaco_2006, Gordeeva_2010, Weir_2013} and at higher propagation speeds in Josephson junction chains~\cite{Pankratov_2012, Soloviev_2013}. Breather excitations are actively investigated in many other areas as well. To cite a few examples, recent publications have examined breathers in cuprate superconductors~\cite{Dienst_2013}, breather wave molecules~\cite{Xu_2019}, breather-type oscillations of the global tectonic shear stress fields~\cite{Zalohar_2020}, matter-wave breathers~\cite{Luo_2020}, and breathers in DNA systems~\cite{Liu_2021}.

The present paper numerically explores the possibility of utilizing magnetic pulses for the generation of travelling breather modes into an LJJ by means of the nonlinear supratransmission~(NST) effect. Nearly two decades ago, it was shown that a nonlinear system subjected to a sinusoidal drive with frequency laying in the forbidden band gap~(FBG) can support energy transmission in the form of solitonic excitations, if the forcing amplitude is high enough. The NST phenomenon, initially discussed in a discrete SG chain by F.~Geniet and J.~Leon~\cite{Geniet_2002, Geniet_2003}, appears to be the result of a generic nonlinear instability~\cite{Leon_2003} and it is nowdays reported in different contexts, e.g., Bragg media~\cite{Leon_2004_1}, wave-guide arrays~\cite{Leon_2004_2,Khomeriki_2004_1,Tog17}, Fermi-Pasta-Ulam model~\cite{Khomeriki_2004_2,Mac18,Tog20}, Klein-Gordon electronic network~\cite{Bodo_2010}, discrete inductance-capacitance electrical line~\cite{Koon_2007}, wave collisions in discrete electrical lattices~\cite{Mot13}, multicomponent non-integrable nonlinear systems~\cite{Vasilescu_2010,Pec21}, stacked Miura-origami metastructure~\cite{Zha20}, and Josephson devices~\cite{Chevriaux_2006, Macias-Diaz_2008, Khomeriki_2009, Mac20}. In the specific case of the LJJ, linear waves, i.e., plasma oscillations, cannot propagate for frequencies lower than the Josephson plasma frequency, $ \omega_p \sim 1 $~THz, hence a pulse with frequency of the same order may be selected to unlock the NST regime. For other examples in which one can successfully use THz radiation to control the macroscopic state in superconductor-based systems, see Refs.~\cite{Schlawin_2017, Schlawin_2019}.

In the presence of dissipation, a current bias, and a thermal noise source, the external pulse's frequency/amplitude space is scanned to find regions where the NST process leads to the exclusive emergence of breather excitations. A thorough analysis of the simulation outcomes reveals that, depending on the shape of the driving field and the values of the main system parameters, vast sets of breather-only frequency/amplitude combinations exist. Moreover, a sort of noise-induced widening of these areas is observed. Interestingly, such an effect is seen to occur in correspondence with a nonmonotonic behavior of the breather-only generation probability, defined below, as a function of the noise amplitude. The fact that a stochastic contribution is included in the picture should not be overlooked. Indeed, although there is a rich literature dealing with environmental effects on the response of Josephson systems~\cite{Castellano_1996, Pankratov_2004, Fedorov_2007, Fedorov_2008, Gordeeva_2008, Fedorov_2009, Guarcello_2013, Valenti_2014, Guarcello_2015_1, Guarcello_2015_Graphene, Guarcello_2016, Guarcello_2017_Graphene, Guarcello_2019_Levy}, NST-related phenomena due to fluctuations have remained largely unaddressed, especially in the realm of the continuous SG model, since the few studies concerning this topic were performed on discrete chains of oscillators, see, e.g., Refs.~\cite{Bodo_2009, Bodo_2013}.

Besides establishing under which conditions the approach works best, the dynamics of the NST-induced breathers is studied by looking at, e.g., the medium's energy and their radiative decay lifetime. The interest in the latter point is stimulated by the lack of information currently available in the literature regarding the influence of a stochastic perturbation term on the breather's annihilation process. The deterministic results presented here are in a good agreement with those predicted by perturbative schemes~\cite{McLaughlin_1978_PRA, Gulevich_2006}, and the estimations performed in the noisy case show a degree of robustness of the breather excitation, with its average persistence time above the fluctuation level being a gradually decreasing function of the noise intensity.

The paper is organized as follows. Section~\ref{Sec2} introduces the model that is assumed to describe the dynamics of a magnetically driven LJJ. Section~\ref{Sec3} presents the numerical results obtained both in a deterministic regime (Sec.~\ref{Sec3a}) and in the presence of thermal fluctuations (Sec.~\ref{Sec3b}). Finally, conclusions are drawn in Sec.~\ref{Sec4}.

\section{Materials and methods}
\label{Sec2}

\begin{figure*}[t!!]
\centering
\includegraphics[width=0.66\textwidth]{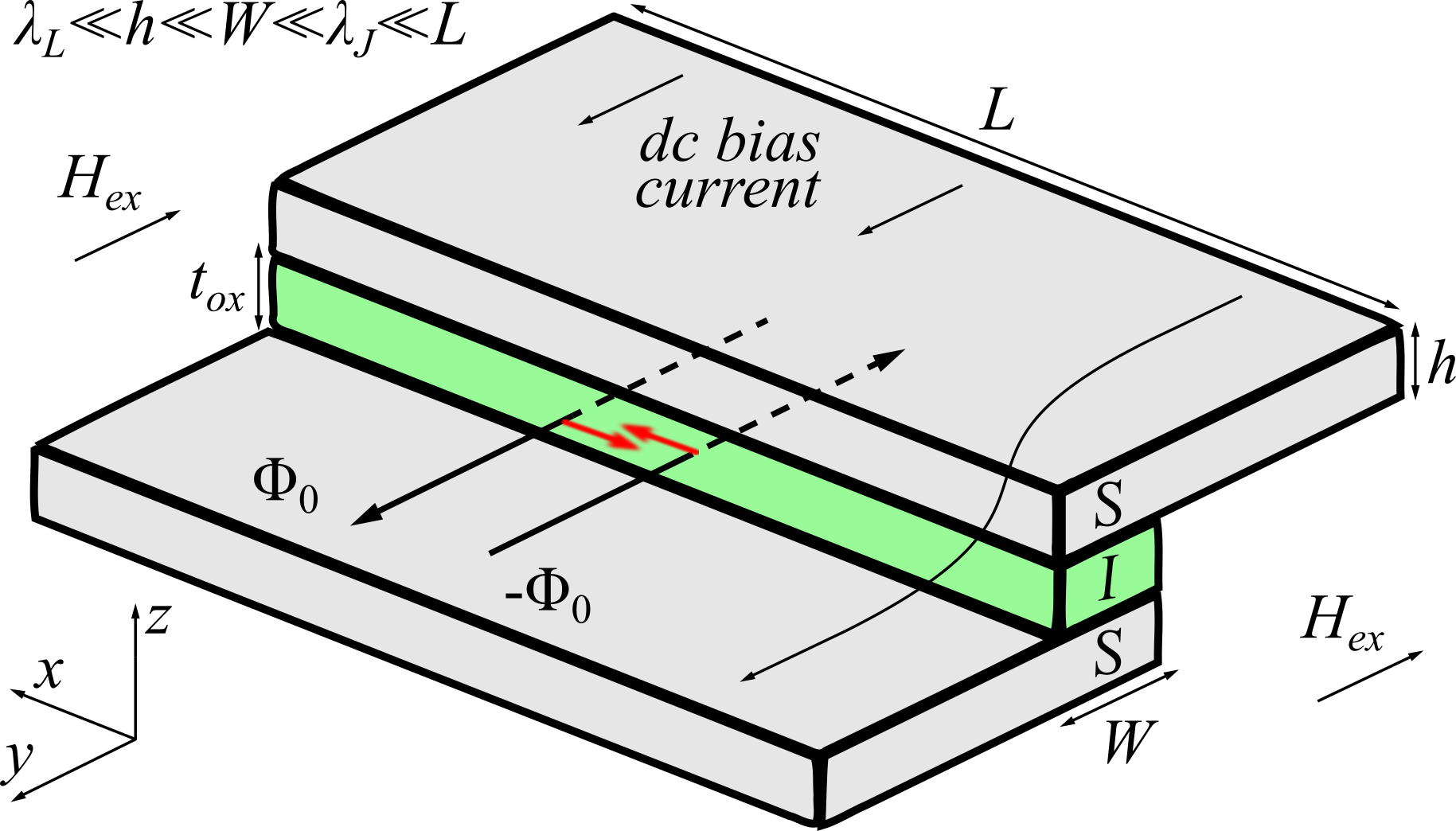}
\caption{Overlap-geometry Josephson tunnel junction (not drawn to scale). The red arrows indicate the attractive interaction between the fluxon and the antifluxon.}
\label{fig:1}
\end{figure*}

It can be shown~\cite{Barone_1982, Castellano_1996} that in an overlap-geometry Josephson tunnel junction, see Fig.~\ref{fig:1}, the dynamics of the phase difference between the pair wave functions of the two superconductors ${ \varphi (x, t) }$ follows the equation
\begin{equation}
\label{eqn:1}
\varphi_{xx} - \varphi_{tt} - \alpha \varphi_{t} = \sin \varphi - \gamma - \gamma_T (x, t) ,
\end{equation}
with the boundary conditions taking into account the magnetic field $ H_{ex} $ applied in the $ y $ direction (see Fig.~\ref{fig:1}) at the edges of the junction
\begin{equation}
\label{eqn:2}
\varphi_{x} (0, t) = \varphi_{x} (l, t) = - \frac{H_{ex} W}{J_c \lambda_J} = \eta .
\end{equation}
In the previous expressions, a subscript notation has been used to denote partial differentiation, i.e., ${ \partial \varphi / \partial t = \varphi_t }$, space is normalized to the Josephson penetration depth ${ \lambda_J = \sqrt{ \Phi_0 / \left( 2 \pi J_c L_P \right)} }$, and time is normalized to the inverse of the Josephson plasma frequency ${ \omega_p = \sqrt{ 2 \pi J_c / \left( \Phi_0 C \right) } }$, where $ J_c $ is the critical value of the Josephson current density, ${ L_P = \mu_0 \left( 2 \lambda_L + t_{ox} \right) / W }$ is the inductance per unit length, and ${ C = \epsilon_r \epsilon_0 W / t_{ox} }$ is the capacitance per unit length. Here, $ \mu_0 $ is the vacuum permeability, $ \lambda_L $ is the London penetration depth, $ t_{ox} $ and $ \epsilon_r $ are, respectively, the thickness and the relative permittivity of the oxide layer, $ W $ is the width of the junction, and $ \epsilon_0 $ is the vacuum permittivity. In Eq.~\eqref{eqn:1}, ${ \alpha = G / \left( \omega_p C \right) }$ is a damping parameter, $ G $ is an effective normal conductance, ${ \gamma = J_b / J_c }$ is the normalized bias current, and ${ \gamma_{T} (x, t) }$ is a Gaussian, zero-average noise source with the autocorrelation function given by
\begin{equation}
\label{eqn:3}
\langle \gamma_{T}(x_1, t_1) \gamma_{T}(x_2, t_2) \rangle = 2 \alpha \Gamma \delta (x_1 - x_2) \delta (t_1 - t_2) .
\end{equation}
Here, ${ \Gamma = 2 e k_B T / \left( \hbar J_c \lambda_J \right) }$ is the noise amplitude, defined in terms of the electron charge $ e $, the Boltzmann constant $ k_B $, the absolute temperature $ T $, and the reduced Planck constant $ \hbar $. Finally, in Eq.~\eqref{eqn:2}, ${ l = L / \lambda_J }$ is the normalized length of the junction and $ \eta $ is a dimensionless quantity proportional to the external magnetic field.

For the sake of completeness, we observe that Eq.~\eqref{eqn:1} does not take into account third-order dissipative effects, i.e., it ignores the term ${ \beta \varphi_{xxt} }$, with ${ \beta = \omega_p L_P / R_P }$, where ${ R_P }$ is related to the scattering of quasiparticles in the surface layers of the two superconductors~\cite{Parmentier_1993}. Also, the spatial gradient of the magnetic field along the junction is neglected, i.e., a contribution ${ \Delta_c H_{x} }$, with ${ \Delta_c }$ being a coupling constant~\cite{Gro92, Parmentier_1993, Kra20}, is not included. The role played by these terms will be investigated elsewhere.

In the absence of perturbations [that is, without the $ \alpha $, $ \gamma $, and ${ \gamma_T (x, t) }$ terms], an infinitely long SG medium sustains solutions of the form
\begin{equation}
\label{eqn:4}
\varphi_{\pm} (x, t) = 4 \arctan \left[ \exp \left( \pm \frac{x - vt}{\sqrt{1 - v^2}} \right) \right] ,
\end{equation} 
known as kinks ($ \varphi_+ $) and antikinks ($ \varphi_- $). In the previous equation, ${ v < 1 }$ is the constant velocity of the travelling wave, which behaves as a relativistic particle due to the structure of the pure SG equation. In the context of LJJs, the limiting velocity ${ \bar{c} = \lambda_J \omega_p = 1 / \sqrt{L_P C} }$ is usually called the Swihart velocity, and it corresponds to the propagation speed of electromagnetic signals in the junction~\cite{Barone_1982}.

A breather is a space-localized, time-oscillating solution of the unperturbed SG equation that reads
\begin{equation}
\label{eqn:5}
\varphi_{b} (x, t) = 4 \arctan \left\lbrace \frac{\sqrt{1 - \omega^2}}{\omega} \frac{ \sin \left[ \frac{\omega \left( t - v_e x \right)}{\sqrt{1 - v_e^2}} \right] } { \cosh \left[ \frac{\sqrt{1 - \omega^2} \left( x - v_e t \right)}{\sqrt{1 - v_e^2}} \right] } \right\rbrace ,
\end{equation}
where ${ \omega < 1 }$ and ${ v_e < 1 }$ are the proper frequency and the envelope velocity of the excitation, respectively. Equation~\eqref{eqn:5} is often regarded as an analytic continuation of the profile describing the kink-antikink collision~\cite{Scott_2003, Dauxois_2006}.

With the aim of determining whether NST constitutes a relevant source of breather modes, Eq.~\eqref{eqn:1} is numerically solved with the initial conditions
\begin{equation}
\label{eqn:6}
\varphi (x, 0) = \arcsin \gamma , \; \; \; \varphi_t (x, 0) = 0 ,
\end{equation}
and the boundary conditions
\begin{equation}
\label{eqn:7}
\varphi_x (0, t) = f (t) , \; \; \; \varphi_x (l, t) = 0 ,
\end{equation}
which represent a junction whose extremity ${ x = 0 }$ is subjected to an external driving ${ f (t) }$. Specifically, the following forcing profile is taken
\begin{equation}
\label{eqn:8}
f (t) = \widetilde{A} (t) \sin \left( \Omega t \right) ,
\end{equation}
with a frequency in the FBG (i.e., ${ \Omega < 1 }$, lower than the plasma value) and, in order to reproduce a meaningful experimental pulse, with Gaussian switching-on/off regimes, i.e.,
\begin{equation}
\label{eqn:9}
\widetilde{A} (t) = \begin{cases}
	A \exp \left[ - \frac{(t - t_{\rm{on}})^2}{2 \sigma_{\rm{on}}^2} \vphantom{\exp \left( - \frac{(t - t_{\rm{off}})^2}{2 \sigma_{\rm{off}}^2} \right)} \right] & t < t_{\rm{on}} \\
	A & t_{\rm{on}} \leq t < t_{\rm{off}} \\
 A \exp \left[ - \frac{(t - t_{\rm{off}})^2}{2 \sigma_{\rm{off}}^2} \right] & t \geq t_{\rm{off}} .
 \end{cases}
\end{equation}
The Gaussian distribution with standard deviation $ \sigma_{\rm{on}} $ in Eq.~\eqref{eqn:9} provides a smoothly increasing signal envelope for ${ t < t_{\rm{on}} }$ (in practice, ${ t_{\rm{on}} = 3 \sigma_{\rm{on}} }$ is chosen). Then, the boundary of the junction is sinusoidally driven until an externally-induced excitation reaches a selected position in the junction, i.e., for ${ t < t_{\rm{off}} }$. If such an event occurs, the driving amplitude is gradually decreased, with the typical time scale $ \sigma_{\rm{off}} $. In this way, the emergence of multiple breathers is largely avoided, and the evolution of the single oscillating mode can be followed.

Since the time $ t_{\rm{off}} $ is not set \textit{a priori}, and it plays a role in the controlled generation of single breather modes, the task of beginning the shutdown of the forcing amplitude should be properly addressed. An externally-induced travelling excitation is assumed to be observed in a designated position ${ x_{\rm{thr}} }$ if the modulus of the phase ${ \left\lvert \varphi \right\rvert }$ at that point exceeds a threshold value $ \varphi_{\rm{thr}} $. This monitoring operation should be performed not too far from the extremity ${ x = 0 }$, but not too close either. In fact, in the former case there would be little to no control over the number of modes populating the junction, while in the latter case the switching process is at risk of being rather intrusive. For the typical value ${ l = 100 }$ employed here, preliminary tests showed that a good choice is ${ x_{\rm{thr}} = 5 }$. As for the phase threshold, ${ \varphi_{\rm{thr}} = 3.5 }$ is fixed to focus the attention on medium-to-large-amplitude excitations and avoid the potential issues related with the distinction between low-energy breathers and plasma-like waves for ${ \Omega \lesssim 1 }$ and ${ A \lesssim 1 }$. However, within reasonable limits, the specific values of $ x_{\rm{thr}} $ and $ \varphi_{\rm{thr}} $ do not significatively alter the outcome of the simulations.

It could be argued that, since in a real situation there is no access to the value of the phase over the LJJ, the above shutdown sequence is somewhat difficult to handle experimentally. While the premise is certainly true, it is worth recognizing that the duration of a pulse which generates a single breather can be readily extracted for any given $ \OA_\star $ combination. Moreover, the same driving field can also be successfully utilized in the neighborhood of $ \OA_\star $, see Sec.~\ref{Sec3}.

\section{Results and discussion}
\label{Sec3}

As stated in Sec.~\ref{Sec1}, one of the aims of the present work is to identify the regions of the $ \OA $ parameter space in which the NST generation mechanism leads to breather excitations only. To this end, the meaning of the expression ``breather-only'' used throughout the paper shall now be operatively defined. Essentially, such a term includes all the cases in which breathers are the only solitons left in the junction from some point onwards, after a possible transitory phase that can involve kinks and antikinks. Conversely, the following situations are not considered to belong to the breather-only class: \emph{i})~at least one bound kink-antikink couple gets separated by the end of the simulation; \emph{ii})~at least one unpaired kink (or antikink) is produced; \emph{iii})~the breather state is not observed at all.

Different techniques can be constructed for the classification of the simulation outcomes. A basic procedure consists in focusing on the motion of a single phase cell~\cite{Geniet_2002, Geniet_2003, phd_Guarcello}, but the results presented here are obtained by means of two new, more reliable criteria. The first strategy is based on the well-known fact that, in the presence of a dissipative perturbation, breathers radiatively decay in time, whereas kinks remain stable. Assuming a simulation time $ t_{\rm{max}} $ sufficiently larger than the expected breather radiative decay lifetime ${ \sim 1 / \alpha }$~\cite{McLaughlin_1978_PRA, Gulevich_2006} (e.g., ${ t_{\rm{max}} \gtrsim 3 / \alpha }$ works well), the condition
\begin{equation}
\label{eqn:10}
\max\limits_{ x \in \left[ 0 , l \right] } \left\lvert \varphi(x, t_{\rm{max}}) \right\rvert \geq 2 \pi
\end{equation}
can only be satisfied if at least one kink (or antikink) is left in the final state of the junction (i.e., at $ t = t_{\rm{max}} $). Conversely, nothing but breather modes are eventually formed in the medium if the threshold $ \varphi_{\rm{thr}} $ is triggered at some instant in time and the relation
\begin{equation}
\label{eqn:11}
\max\limits_{ x \in \left[ 0 , l \right] } \left\lvert \varphi(x, t_{\rm{max}}) \right\rvert < 2 \pi
\end{equation}
holds~\footnote[1]{In practice, due to possible reflection effects at the boundaries of the system, the values ${ \left\lvert \varphi (0, t) \right\rvert }$ and ${ \left\lvert \varphi (l, t) \right\rvert }$ require additional monitoring.}. In the following, Eq.~\eqref{eqn:11} will be referred to as the \textit{phase criterion}. 

The other approach takes advantage of the practically null average voltage drop across the junction produced by breathers. Given the structure of the second Josephson relation~\cite{Barone_1982}, it is possible to define the (normalized) time-averaged voltage
\begin{equation}
\label{eqn:12}
\left\langle V \right\rangle = \Big\langle \int_0^l \varphi_t dx \Big\rangle .
\end{equation}
Through extensive preliminary tests (the usual simulation parameters are listed in Sec.~\ref{Sec3a}), it was seen that the exclusive emergence of breather modes takes place if $ \varphi_{\rm{thr}} $ is reached at some time and the condition
\begin{equation}
\label{eqn:13}
\left\lvert \langle V \rangle \right\rvert \lesssim 0.05
\end{equation}
is verified~\footnote[2]{See the first footnote.}, otherwise at least one kink (or antikink) is left in the medium at ${ t = t_{\rm{max}} }$. Equation~\eqref{eqn:13} will be indicated hereafter as the \textit{voltage criterion}.

As will be discussed below, these two alternatives are almost equally dependable. On rare occasions, mainly due to pronounced kink-antikink-like transients, the voltage criterion was found to be inaccurate, however it has the advantage of relying on an intuitive, measurable physical quantity, and it could be useful for an experimental test of this method.

It should be emphasized that the presence of multiple breathers does not constitute an issue for the detection techniques introduced above. The choice of working with single breathers is made in order to investigate their dynamics, without the complications arising from the interaction with other emitted modes. Moreover, the forcing envelope defined by Eq.~\eqref{eqn:9} cannot guarantee the generation of single breathers in every possible situation. In few cases, exclusively near the phonon band (${ \Omega \lesssim 1 }$), the generation of two (or three) very close low-energy breathers may still be encountered. Hence, developing additional tests to efficiently identify these events is important to avoid a blind application of concepts specifically designed for single excitations.

\subsection{Deterministic analysis}
\label{Sec3a}

\begin{figure*}[t!!]
\centering
\includegraphics[width=\textwidth]{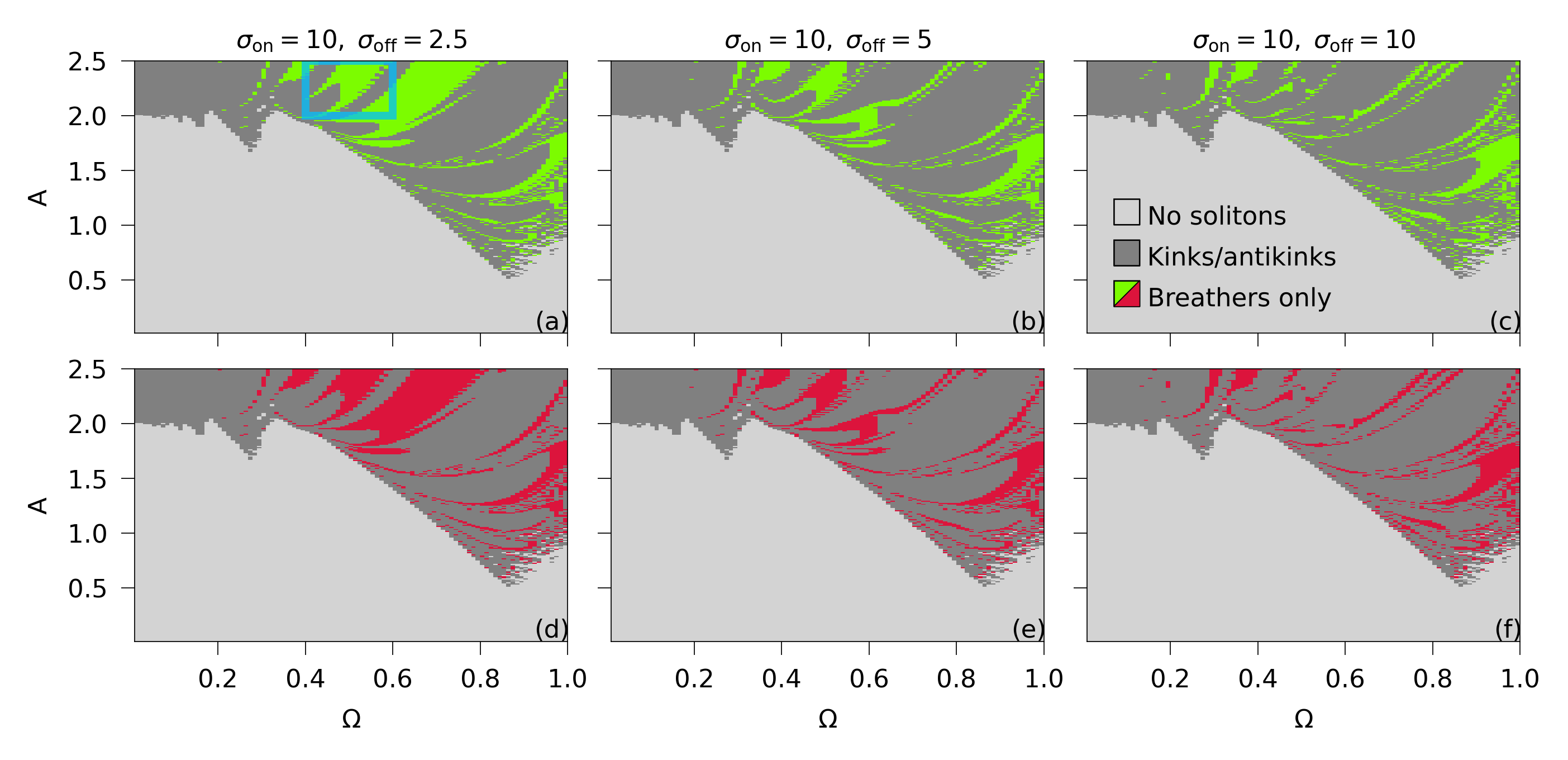}
\caption{Bifurcation diagrams in the $ \OA $ plane. The upper panels are obtained by means of the phase criterion, see Eq.~\eqref{eqn:11}, while the lower ones result from the application of the voltage criterion, see Eq.~\eqref{eqn:13}. The light gray color indicates $ \OA $ couples for which no excitations of clear solitonic nature are detected in the medium, the dark gray color corresponds to regions with at least one kink (or antikink) left in the junction at ${ t = t_{\rm{max}} }$, and the green (red) color is exclusively associated with breathers according to the phase (voltage) criterion. Here, ${ l = 100 }$, ${ t_{\rm{max}} = 200 }$, ${ \alpha = 0.02 }$, ${ \gamma = 0 }$, and ${ \sigma_{\rm{on}} = 10 }$ are fixed, whereas $ \sigma_{\rm{off}} $ is varied. In particular, ${ \sigma_{\rm{off}} = 2.5 }$ in panels (a) and (d), ${ \sigma_{\rm{off}} = 5 }$ in (b) and (e), and ${ \sigma_{\rm{off}} = 10 }$~in~(c)~and~(f). The blue box in panel (a) delimits the $ \OA $ subsection ${ \left[ 0.4 , 0.6 \right] \times \left[ 2 , 2.5 \right] }$.}
\label{fig:2}
\end{figure*}
In this subsection, the stochastic term ${ \gamma_T (x, t) }$ is not taken into account. The resulting partial differential equation is numerically integrated through a Fortran implementation of an implicit finite-difference scheme~\cite{Ames_1977, Press_1992}, with discretization steps ${ \Delta x = \Delta t = 0.005 }$. The typical junction length is ${ l = 100 }$, for which reflection effects at ${ x = l }$ are generally negligible. The observation time is usually set to ${ t_{\rm{max}} = 200 }$ in order to follow the entire evolution of the emitted breathers. Also, the $ \OA $ parameter space is explored here by varying the driving frequency $ \Omega $ in the range ${ [0, 1] }$ and the amplitude $ A $ in the range ${ [0, 2.5] }$, with increments ${ \Delta \Omega = \Delta A = 0.01 }$.

\begin{figure*}[t!!]
\centering
\includegraphics[width=0.75\textwidth]{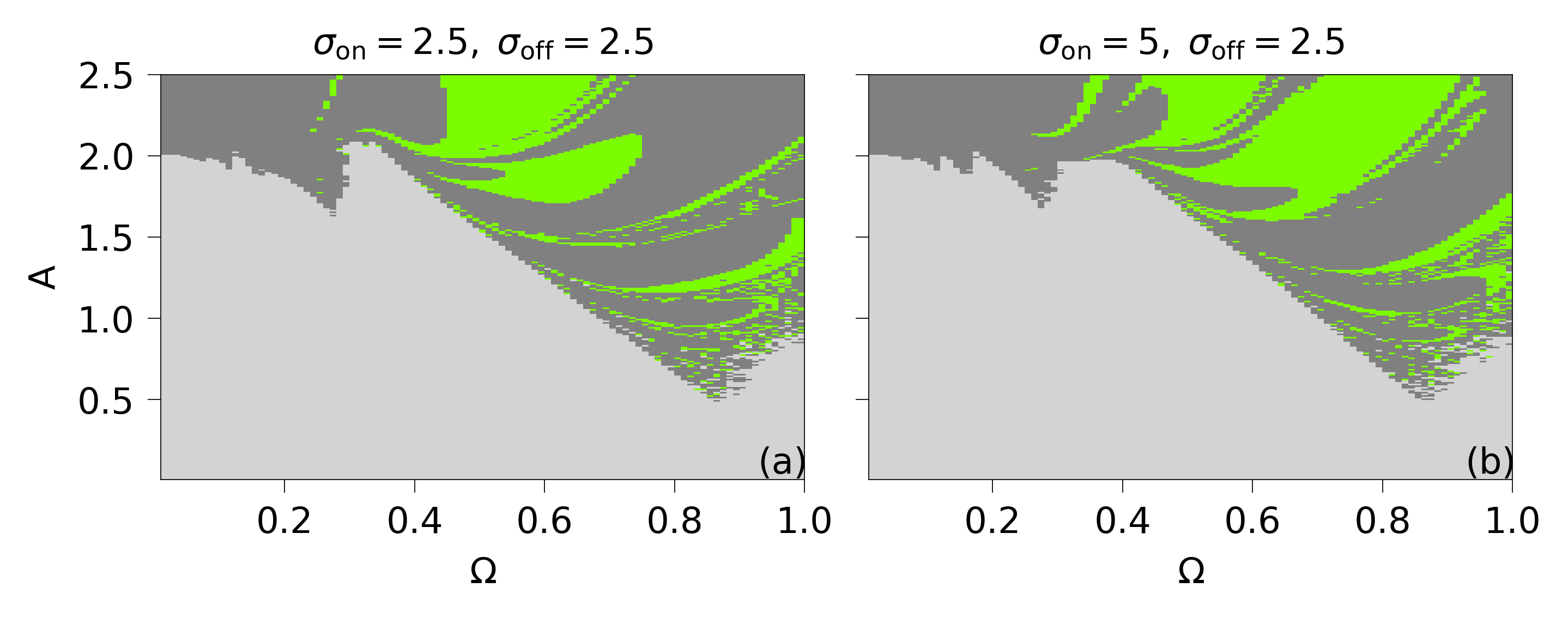}
\caption{Bifurcation diagrams in the $ \OA $ plane. The light gray color indicates $ \OA $ couples for which no excitations of clear solitonic nature are detected in the medium, the dark gray color corresponds to regions with at least one kink (or antikink) left in the junction at ${ t = t_{\rm{max}} }$, and the green color is exclusively associated with breathers according to the phase criterion, see Eq.~\eqref{eqn:11}. Here, ${ l = 100 }$, ${ t_{\rm{max}} = 200 }$, ${ \alpha = 0.02 }$, ${ \gamma = 0 }$, and ${ \sigma_{\rm{off}} = 2.5 }$ are fixed, whereas $ \sigma_{\rm{on}} $ is varied. In particular, ${ \sigma_{\rm{on}} = 2.5 }$ in panel (a) and ${ \sigma_{\rm{on}} = 5 }$ in panel (b).}
\label{fig:3}
\end{figure*}
For the first set of simulations, the damping coefficient is ${ \alpha = 0.02 }$~\cite{Golovchanskiy_2017}, the bias current is ${ \gamma = 0 }$, the width of the increasing Gaussian tail is ${ \sigma_{\rm{on}} = 10 }$, and ${ \sigma_{\rm{off}} = 2.5, 5, 10 }$ to investigate the effect of the duration of the drive's switching-off regime on the generation process. Going beyond the identification of the NST activation threshold~\cite{Geniet_2002, Geniet_2003}, Fig.~\ref{fig:2} presents the results in the form of refined bifurcation diagrams. More specifically, the upper panels are obtained by means of the phase criterion [Eq.~\eqref{eqn:11}], while the lower ones come from the application of the voltage criterion [Eq.~\eqref{eqn:13}]. In these plots: \emph{i})~the light gray color indicates $ \OA $ couples for which no excitations of clear solitonic nature are detected in the medium, i.e., satisfying ${ \varphi (x_{\rm{thr}}, t \leq t_{\rm{max}}) < \varphi_{\rm{thr}} }$; \emph{ii})~the dark gray color corresponds to regions with at least one kink (or antikink) left in the junction at ${ t = t_{\rm{max}} }$; \emph{iii})~the green (red) color is exclusively associated with breathers according to the phase (voltage) criterion. The presence of significant breather-only areas is evident, especially for ${ \sigma_{\rm{off}} = 2.5 }$. The gradual disappearance of those regions for higher values of ${ \sigma_{\rm{off}} }$ is comprehensible, since additional kinks (and/or antikinks) are more likely generated after a breather if the forcing signal takes a longer time to stop injecting energy into the medium. Another noteworthy aspect of Fig.~\ref{fig:2} is that the two detection strategies yield very close bifurcation diagrams. Since this holds for all the used parameter sets, reporting the results produced by both Eq.~\eqref{eqn:11} and Eq.~\eqref{eqn:13} in any case is not necessary, so the rest of the paper will focus on the phase criterion.

Figure~\ref{fig:3} illustrates the diagrams computed with ${ \sigma_{\rm{on}} = 2.5 }$ and ${ \sigma_{\rm{on}} = 5 }$, fixing ${ \alpha = 0.02 }$, ${ \gamma = 0 }$, and ${ \sigma_{\rm{off}} = 2.5 }$. The two panels, together with Fig.~\ref{fig:2}(a), seem to be characterized by a similar structure, but the variation of $ \sigma_{\rm{on}} $ unquestionably affects the precise shape and localization of the zones of interest. In the following, ${ \sigma_{\rm{on}} = 10 }$ and ${ \sigma_{\rm{off}} = 2.5 }$ are set, however the following procedures lead to perfectly equivalent results for the other values of these parameters.

\begin{figure*}[t!!]
\centering
\includegraphics[width=0.75\textwidth]{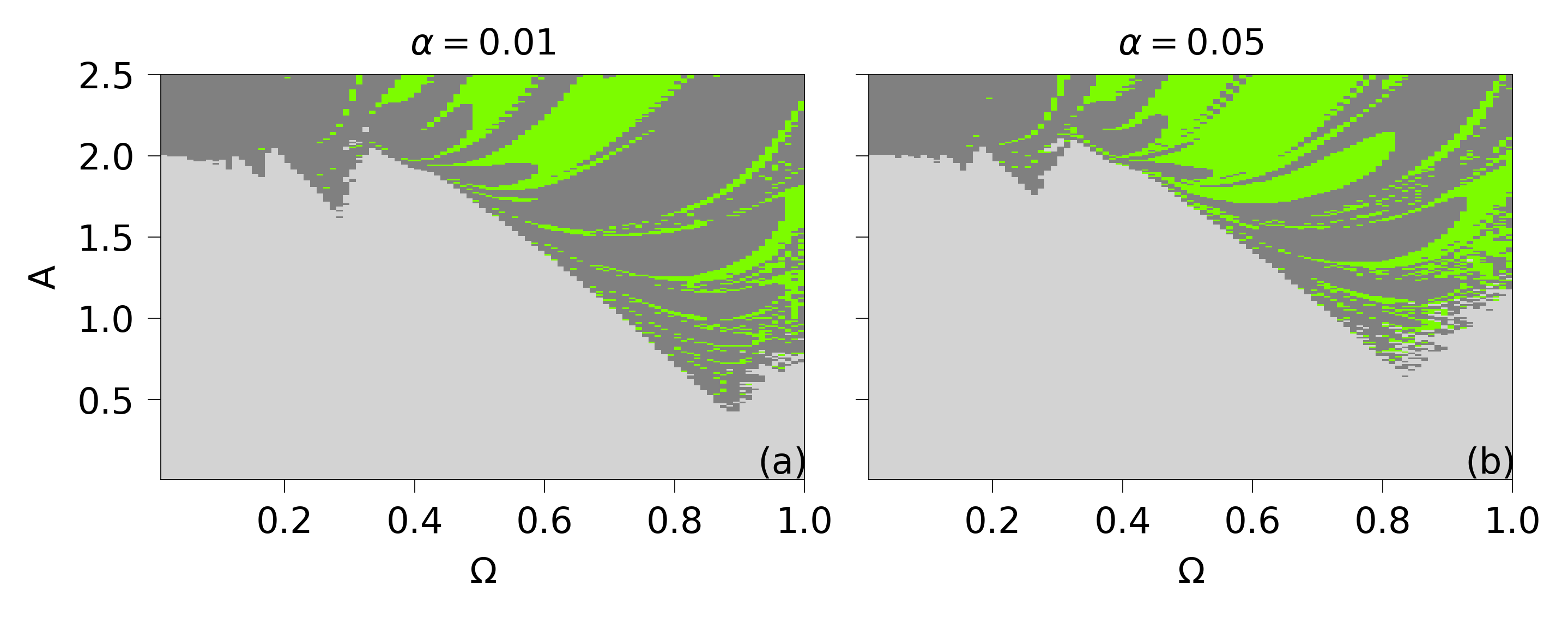}
\caption{Bifurcation diagrams in the $ \OA $ plane. The light gray color indicates $ \OA $ couples for which no excitations of clear solitonic nature are detected in the medium, the dark gray color corresponds to regions with at least one kink (or antikink) left in the junction at ${ t = t_{\rm{max}} }$, and the green color is exclusively associated with breathers according to the phase criterion, see Eq.~\eqref{eqn:11}. Here, ${ \gamma = 0 }$, ${ \sigma_{\rm{on}} = 10 }$, and ${ \sigma_{\rm{off}} = 2.5 }$ are fixed, whereas ${ \alpha }$ is varied. In particular, ${ \alpha = 0.01 }$, ${ l = 200 }$, and ${ t_{\rm{max}} = 300 }$ in panel (a) and ${ \alpha = 0.05 }$, ${ l = 100 }$, and ${ t_{\rm{max}} = 200 }$ in panel (b).}
\label{fig:4}
\end{figure*}
The influence of the dissipation is now studied by varying the damping parameter $ \alpha $ in the range ${ \left[ 0.01, 0.05 \right] }$. Moreover, because for $ \alpha = 0.01 $ breathers are expected to last longer, for this specific value the length of the junction and the observation time are set to $ l = 200 $ and $ t_{\rm{max}} = 300 $, respectively.

The bifurcation diagrams obtained for ${ \alpha = 0.01 }$ and ${ \alpha = 0.05 }$, see Fig.~\ref{fig:4}, show that the extension of the breather-only regions grows with the damping coefficient. This is reasonable, since possible NST-induced kink-antikink pairs are more likely to relax into the oscillatory bound state for greater values of $ \alpha $~\cite{Scott_2003, Dauxois_2006}.

\begin{figure*}[t!!]
\centering
\includegraphics[width=\textwidth]{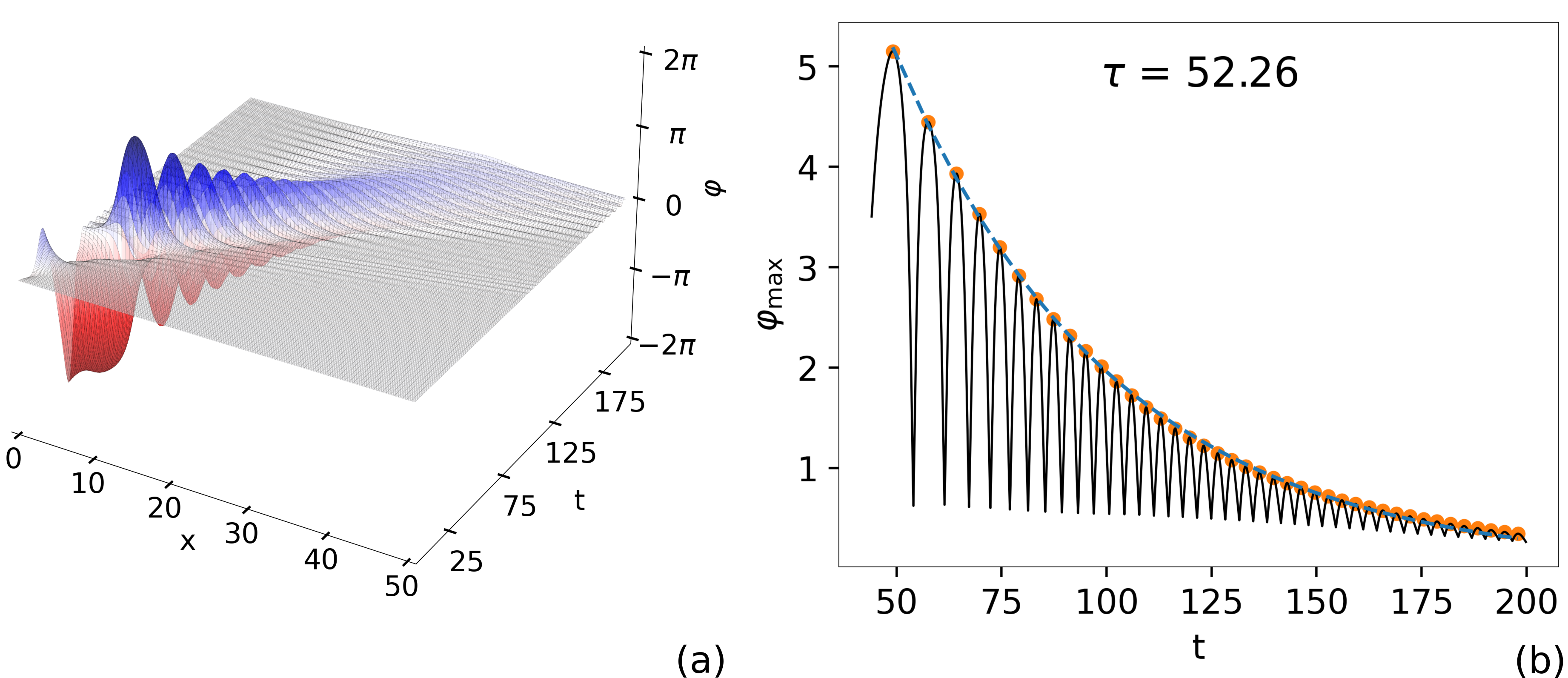}
\caption{Panel~(a): Time evolution of a single NST-induced travelling breather in the presence of a dissipative perturbation. Panel~(b): Time evolution of ${ \varphi_{\rm{max}} }$ (black curve), that is the amplitude of the decaying breather shown in panel~(a), recorded from the time instant at which the value $ \varphi_{\rm{thr}} $ is reached in $ x_{\rm{thr}} $, i.e., for ${ t \geq t_{\rm{off}} }$ [see Eq.~\eqref{eqn:14}]. Each peak, see the orange circles, corresponds to a half-cycle of the breather oscillation. The exponential curve that best fits the detected peaks is represented by a blue dashed line, and the inverse of its decay rate is ${ \tau = 52.26 }$. The parameter values of the simulation are ${ l = 100 }$, ${ t_{\rm{max}} = 200 }$, ${ \alpha = 0.02 }$, ${ \gamma = 0 }$, ${ \Omega = 0.25 }$, ${ A = 2.09 }$, ${ \sigma_{\rm{on}} = 10 }$, and ${ \sigma_{\rm{off}} = 2.5 }$.}
\label{fig:5}
\end{figure*}
For the single-breather cases identified by the phase criterion, a radiative decay lifetime $ \tau $ can be estimated by recording the maximum value of ${ \left\lvert \varphi (x, t) \right\rvert }$ along the junction~\footnote[3]{In practice, due to possible reflection effects at the boundaries of the system, a subinterval ${ \left[ \delta , l - \delta \right] }$, with ${ 0 < \delta \ll l }$, is chosen.}
\begin{equation}
\label{eqn:14}
\varphi_{\rm{max}}(t) = \max\limits_{x \in \left[ 0 , l \right] } \left\lvert \varphi (x, t) \right\rvert
\end{equation}
from the time instant at which ${ \varphi_{\rm{thr}} }$ is triggered by some excitation, i.e., for ${ t \geq t_{\rm{off}} }$. In fact, the damped oscillations of this quantity correspond to the breather's decaying amplitude~\footnote[4]{Since the quantity in Eq.~\eqref{eqn:14} does not automatically adapt to the presence of multiple breathers, it is important to avoid utilizing it in the latter case. Fortunately, the presence of multiple breather modes is not particularly hard to detect. For example, if one correctly manages the distorted oscillating profile of the travelling breather, a simple strategy consists in monitoring the number of wave packets that pass through a selected position relatively near to ${ x = 0 }$.}. More precisely, as shown in Fig.~\ref{fig:5}, obtained for ${ \alpha = 0.02 }$, ${ \Omega = 0.25 }$, and ${ A = 2.09 }$, since the local maxima of ${ \varphi_{\rm{max}}(t) }$ follow an exponential trend [see the orange circles in Fig.~\ref{fig:5}(b)], a standard fitting procedure is applied to these points, defining the radiative decay lifetime ${ \tau }$ as the inverse of the exponential decay rate.

\sloppy In some cases, an initial time interval ${ t_{ 2 \pi } }$ such that ${ \varphi_{\rm{max}}( t_{\rm{off}} \leq t < t_{\rm{off}} + t_{ 2 \pi }) \sim 2 \pi }$ is observed, and its duration can be significantly higher than each of the subsequent oscillations of exponentially decreasing amplitude. This transient phase profile can supposedly be ascribed to a kink-antikink-like state, e.g., it may correspond to the half-cycle of oscillation of a very low frequency breather. To further address the point, the characteristic time ${ t_{ 2 \pi } }$ is recorded whenever present. In light of the above, it is convenient to employ a fitting profile of the form
\begin{equation}
\label{eqn:21}
g (t) = \begin{cases}
	B & t_{\rm{off}} \leq t < t_{\rm{off}} + t_{ 2 \pi } \\
	B \cdot {\rm{e}}^{ - \frac{ t - \left( t_{\rm{off}} + t_{ 2 \pi } \right) }{\tau} } & t \geq t_{\rm{off}} + t_{ 2 \pi } ,
 \end{cases}
\end{equation}
with ${ B \sim 2 \pi }$.

\begin{figure*}[t!!]
\centering
\includegraphics[width=0.75\textwidth]{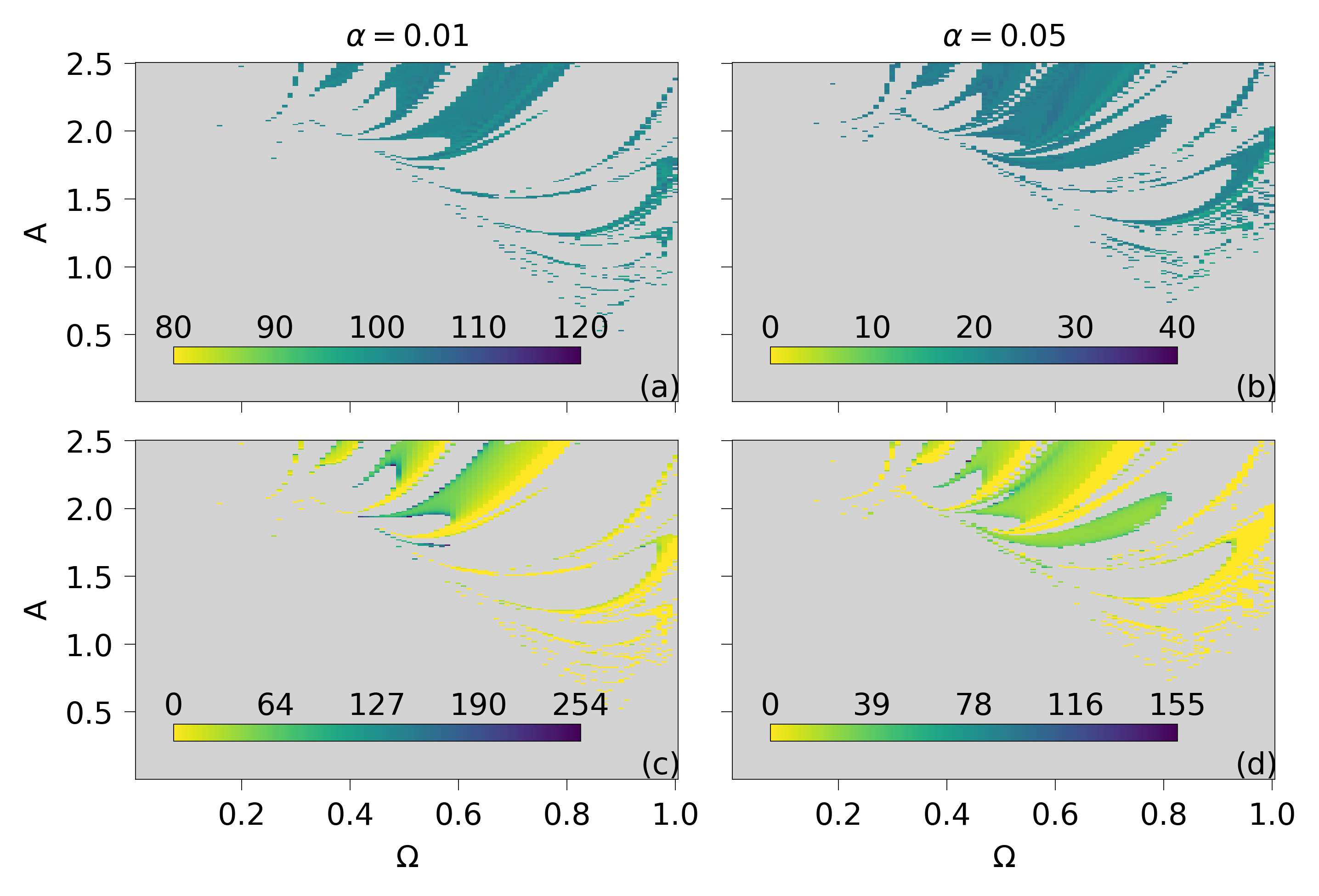}
\caption{Breather radiative decay lifetime $ \tau $ (upper panels) and ${ t_{2 \pi} }$ (lower panels) as a function of the driving frequency $ \Omega $ and amplitude $ A $. The light gray color corresponds to $ \OA $ couples for which these quantities cannot be estimated. Here, ${ \gamma = 0 }$, ${ \sigma_{\rm{on}} = 10 }$, and ${ \sigma_{\rm{off}} = 2.5 }$ are fixed, while $ \alpha $ is varied. In particular, ${ \alpha = 0.01 }$, ${ l = 200 }$, and ${ t_{\rm{max}} = 300 }$ in panels (a) and (c) and ${ \alpha = 0.05 }$, ${ l = 100 }$, and ${ t_{\rm{max}} = 200 }$ in panels (b) and (d).}
\label{fig:6}
\end{figure*}
%
Systematic evaluations of the fitting parameter $ \tau $, for the single-breather cases detected by means of the phase criterion, are displayed in the upper panels of Fig.~\ref{fig:6}. It should be stressed that this estimation is only performed on the time series ${ \varphi_{\rm{max}} (t) }$ which contain at least five peaks of oscillation. Also, some distorted decay signals due to breathers stuck at the boundary ${ x = 0 }$ were discarded.

\begin{figure*}[t!!]
\centering
\includegraphics[width=0.5\textwidth]{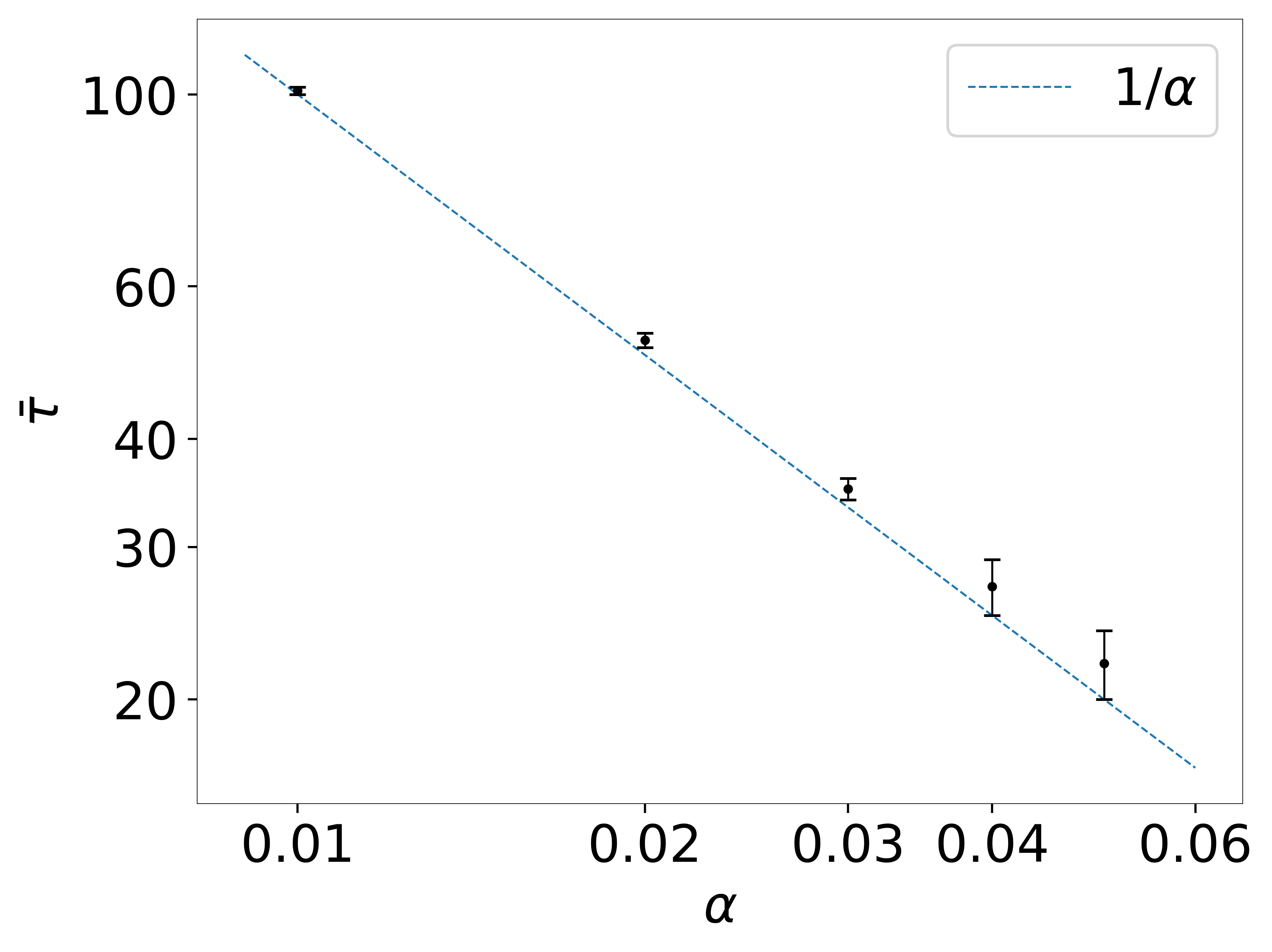}
\caption{Log-log plot of the breather lifetime averaged over the $ \OA $ parameter space, ${ \bar{\tau} }$, and its standard deviation, ${ \sigma }$, extracted from the simulations (black points). The dashed blue line represents the perturbative prediction.}
\label{fig:7}
\end{figure*}

The quantity $ \tau $ is seen to decrease with $ \alpha $, almost independently of the $ \OA $ combination, in accordance with the perturbative prediction, i.e., ${ \tau \sim 1 / \alpha }$. This fact can be better visualized by plotting the average value over the $ \OA $ space, ${ \bar{\tau} }$, and its standard deviation, ${ \sigma }$, as a function of the damping coefficient. As illustrated by the log-log graph in Fig.~\ref{fig:7}, the average simulated lifetime appears to be close (slightly over) the value ${ 1 / \alpha }$.

The lower panels of Fig.~\ref{fig:6} show the ${ t_{ 2 \pi} }$ time intervals measured for the single-breather cases identified in Fig.~\ref{fig:4}. In contrast to the diagrams concerning ${ \tau }$, these highlight a sort of structure internal to each breather-only zone, which is not necessarily shared with the neighboring ones. Furthermore, an overall decrease of ${ t_{ 2 \pi} }$ for higher values of the damping coefficient is observed.

\begin{figure*}[t!!]
\centering
\includegraphics[width=0.75\textwidth]{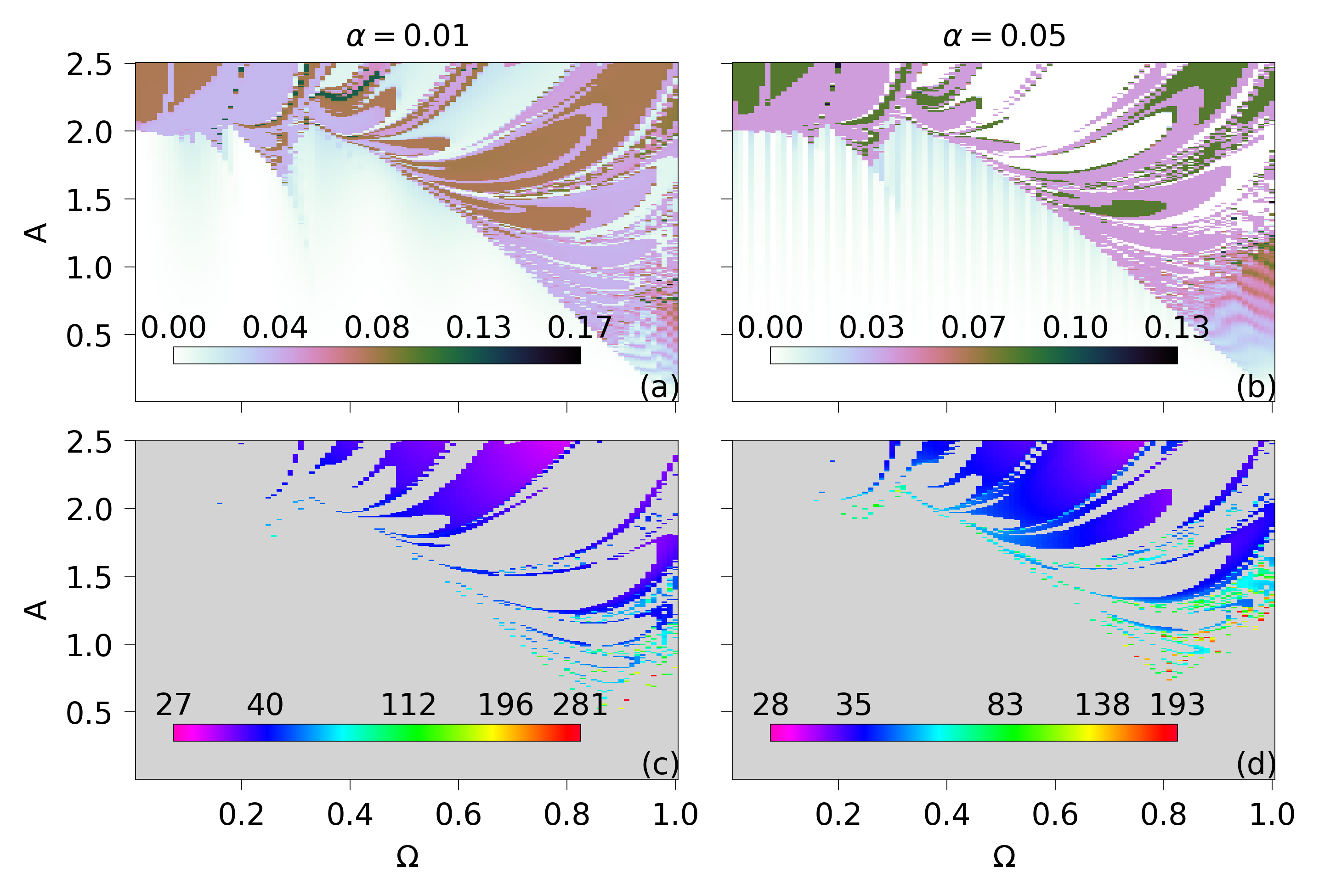}
\caption{Numerically evaluated energy (upper panels) and excitation time $ t_{\rm{off}} $ (lower panels) as a function of the driving frequency $ \Omega $ and amplitude $ A $. In the lower panels, the light gray color corresponds to $ \OA $ couples that do not belong to the breather-only areas according to the phase criterion, see Eq.~\eqref{eqn:11}. Also, a power-law scale is employed for visualization purposes. Here, ${ \gamma = 0 }$, ${ \sigma_{\rm{on}} = 10 }$, and ${ \sigma_{\rm{off}} = 2.5 }$ are fixed, while $ \alpha $ is varied. In particular, ${ \alpha = 0.01 }$, ${ l = 200 }$, and ${ t_{\rm{max}} = 300 }$ in panels (a) and (c) and ${ \alpha = 0.05 }$, ${ l = 100 }$, and ${ t_{\rm{max}} = 200 }$ in panels (b) and (d).}
\label{fig:8}
\end{figure*}

In the NST context, another typically interesting quantity to look at is the medium's energy. Ignoring the ${ \gamma_T (x, t) }$ term, the space-discretized version of Eq.~\eqref{eqn:1} is
\begin{equation}
\label{eqn:15}
c^2 (\varphi_{n + 1} - 2 \varphi_n + \varphi_{n - 1}) - \ddot{\varphi}_n - \alpha \dot{\varphi}_n = \sin \varphi_n - \gamma ,
\end{equation}
in which ${ n = 1, ..., \infty }$, ${ c = 1 / \Delta x }$, and Newton's notation has been used to indicate time differentiation. If the following expression is kept for the energy density
\begin{equation}
\label{eqn:16}
H_n = \frac{1}{2} \dot{\varphi}_n^2 + \frac{c^2}{2} (\varphi_{n + 1} - \varphi_{n})^2 + 1 - \cos \varphi_n , 
\end{equation}
by taking its time derivative, with the aid of Eq.~\eqref{eqn:15}, one gets
\begin{equation}
\label{eqn:17}
\dot{H}_n = - \alpha \dot{\varphi}_n^2 + \gamma \dot{\varphi}_n + J_n - J_{n + 1} ,
\end{equation}
where the current is defined as
\begin{equation}
\label{eqn:18}
J_n = - c^2 \dot{\varphi}_n (\varphi_n - \varphi_{n - 1}) .
\end{equation}
In analogy with Refs.~\cite{Geniet_2002, Geniet_2003}, the total energy is obtained by inserting a boundary energy contribution, i.e.,
\begin{equation}
\label{eqn:19}
E = \sum_{n = 1}^{\infty} H_n + \frac{c^2}{2} (\varphi_1 - \varphi_0)^2 .
\end{equation}
Assuming that ${ J_n \to 0 }$ as ${ n \to \infty }$, a direct evaluation of the terms in the previous expression leads to
\begin{equation}
\label{eqn:20}
E = \int_0^{t_{\rm{max}}} \left[ - \alpha \sum_{n = 1}^\infty \dot{\varphi}_n^2 + \gamma \sum_{n = 1}^\infty \dot{\varphi}_n + c^2 \dot{\varphi_0} (\varphi_0 - \varphi_1) \right] dt .
\end{equation}
Numerical computations of the quantity ${ E \; {\Delta x}^2 }$ are displayed in the upper panels of Fig.~\ref{fig:8} for the same values of $ \alpha $ used in Fig.~\ref{fig:4}, clearly exhibiting the sudden energy flow associated with the NST process. Since ${ t_{\rm{max}} \gtrsim 3 / \alpha }$, by the end of the simulations most of the breathers have died out. Hence, breather-only zones can be recognized in these maps by their low ${ E \; {\Delta x}^2 }$ values. Also, one may notice the presence of quasi-periodic patterns below the NST threshold, i.e., in the region of the $ \OA $ plane where it should not be possible to transmit energy to the medium. Simple considerations on the structure of Eq.~\eqref{eqn:20} suggest that this effect stems from the dependence of its last term on the simulation time~\footnote[5]{This can be confirmed, e.g., by computing an energy map for different values of ${ t_{\rm{max}} }$ (not shown here).}.

In support of the closing remark of Sec.~\ref{Sec2}, the lower panels of Fig.~\ref{fig:8} display the value of the excitation time $ t_{\rm{off}} $, i.e., the time at which $ \varphi_{\rm{thr}} $ is triggered in $ x_{\rm{thr}} $, as a function of the driving frequency $ \Omega $ and amplitude $ A $. The light gray color corresponds to $ \OA $ couples that do not belong to the breather-only areas according to the phase criterion, see Eq.~\eqref{eqn:11}. Also, a power-law scale is employed for visualization purposes. The distribution of the excitation times in the parameter space appears to be smooth over the zones of interest, making it possible to identify a typical pulse duration suitable for vast $ \OA $ regions~\footnote[6]{As one may guess, a change in the time scale of the switching-on regime of the driving signal (i.e., $ \sigma_{\rm{on}} $) alters the typical value of the excitation time $ t_{\rm{off}} $. Although not explicitly addressed in the paper, very similar distributions of the quantity ${ t_{\rm{off}} \OA }$ are found for different values of $ \sigma_{\rm{on}} $.}. Indeed, further testing has shown that qualitatively analogous breather-only areas can also be obtained with fixed-duration forcing envelopes, i.e., with pulses that do not automatically shut down after an externally-induced excitation is detected.

\begin{figure*}[t!!]
\centering
\includegraphics[width=0.75\textwidth]{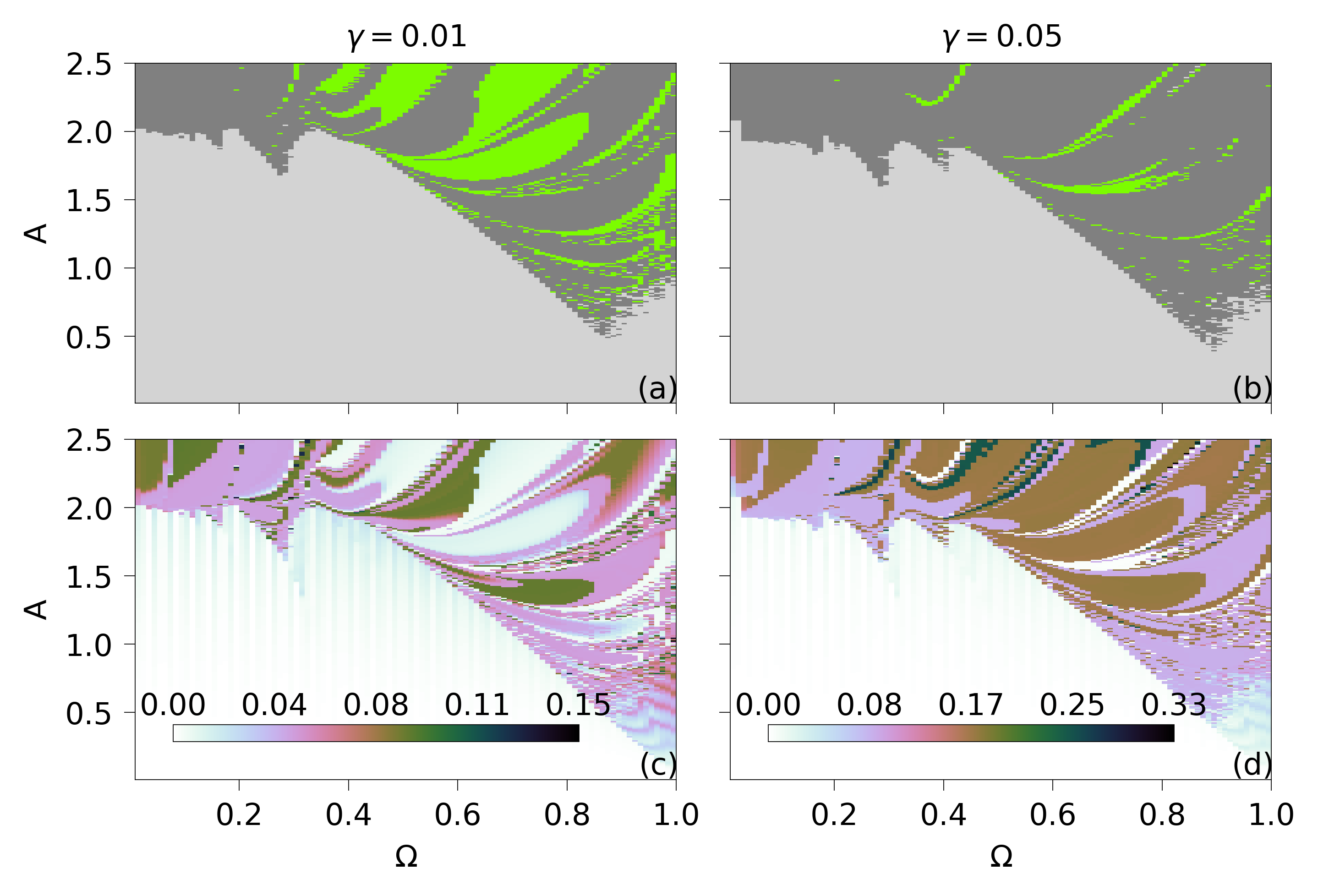}
\caption{Upper panels: Bifurcation diagrams in the $ \OA $ plane. The light gray color indicates $ \OA $ couples for which no excitations of clear solitonic nature are detected in the medium, the dark gray color corresponds to regions with at least one kink (or antikink) left in the junction at ${ t = t_{\rm{max}} }$, and the green color is exclusively associated with breathers according to the phase criterion, see Eq.~\eqref{eqn:11}. Lower panels: Numerically evaluated energy as a function of the driving frequency $ \Omega $ and amplitude $ A $. Here, ${ l = 100 }$, ${ t_{\rm{max}} = 200 }$, ${ \alpha = 0.02 }$, ${ \sigma_{\rm{on}} = 10 }$, and ${ \sigma_{\rm{off}} = 2.5 }$ are fixed, whereas $ \gamma $ is varied. In particular, ${ \gamma = 0.01 }$ in panels (a) and (c) and ${ \gamma = 0.05 }$ in panels (b) and (d).}
\label{fig:9}
\end{figure*}

Finally, two $ \OA $ bifurcation diagrams with the damping parameter set to ${ \alpha = 0.02 }$ and the bias current assuming the values ${ \gamma = 0.01 }$ and ${ \gamma = 0.05 }$ are presented in the upper panels of Fig.~\ref{fig:9}. Here, the breather-only zones are seen to progressively vanish as the bias current gets stronger, since the $ \gamma $ perturbation term tends to tear the kink-antikink couple apart~\cite{Gulevich_2012}. Interestingly, Fig.~\ref{fig:9}(a) also shows that, for ${ \gamma \lesssim \alpha }$, some breather-only areas are expanded, i.e., small current values can support the formation of breather modes.

According to the estimation procedure discussed above, no appreciable modifications to the lifetime $ \tau $ occur due to the $ \gamma $ term, which indicates that the bias current is not able to effectively transfer energy to the oscillatory bound state, and the breather eventually dies out as it would happen for ${ \gamma = 0 }$. This fact is confirmed by the energy diagrams, i.e., the lower panels of Fig.~\ref{fig:9}, where the breather-only regions are still clearly distinguishable by their significantly low ${ E \; {\Delta x}^2 }$ values, while gradually more energy is seen to enter the surrounding areas, above the NST threshold, as $ \gamma $ rises. Considerations similar to those made earlier apply to the maps of the excitation time $ t_{\rm{off}} $, which seem to maintain their macroscopic structure independently of the value of the bias current.

\subsection{Noisy dynamics}
\label{Sec3b}

\begin{figure*}[t!!]
\centering
\includegraphics[width=\textwidth]{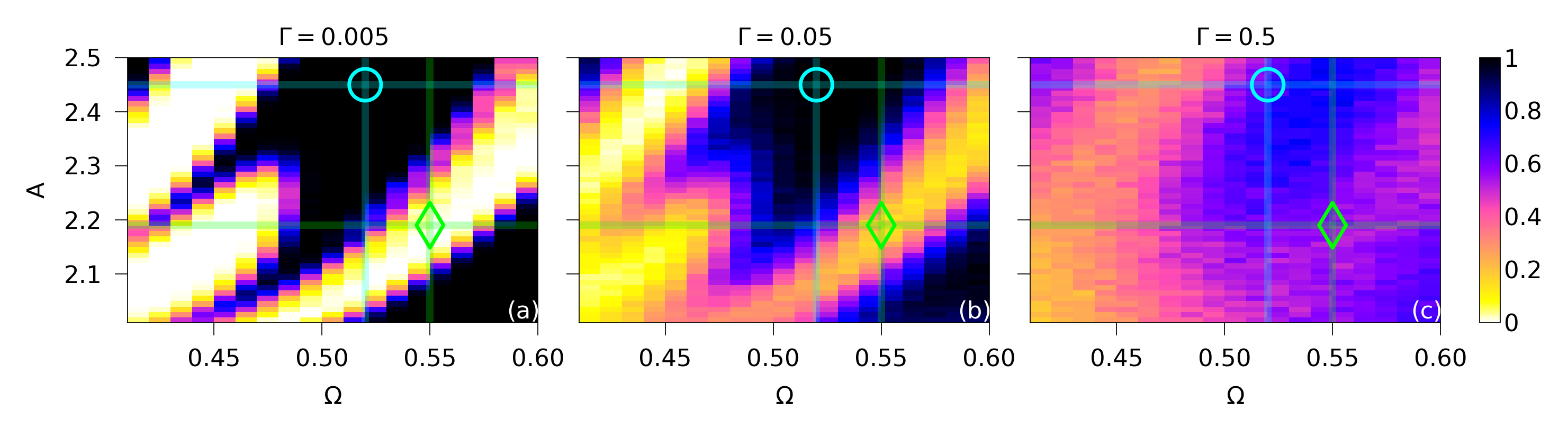}
\caption{Breather-only generation probability estimated over $ N = 512 $ realizations according to the phase criterion, see Eq.~\eqref{eqn:11}. Here, ${ l = 100 }$, ${ \alpha = 0.02 }$, ${ \gamma = 0 }$, ${ \sigma_{\rm{on}} = 10 }$, and ${ \sigma_{\rm{off}} = 2.5 }$ are fixed, while $ \Gamma $ is varied. In particular, ${ \Gamma = 0.005 }$ and ${ t_{\rm{max}} = 150 }$ in panel (a), ${ \Gamma = 0.05 }$ and ${ t_{\rm{max}} = 150 }$ in panel (b), and ${ \Gamma = 0.5 }$ and ${ t_{\rm{max}} = 200 }$ in panel (c). The green diamonds (cyan circles) identify the combination ${ \Omega = 0.55 }$ and ${ A = 2.19 }$ (${ \Omega = 0.52 }$ and ${ A = 2.45 }$).}
\label{fig:11}
\end{figure*}
Having documented the observation of large breather-only regions in the $ \OA $ parameter space, a crucial question that remains to be answered is whether these structures are robust enough to survive in a noisy environment. To this end, the randomly fluctuating current ${ \gamma_{T} (x, t) }$ is now included (see Ref.~\cite{Tuckwell_2016} for discussions concerning the term's numerical approximation). Here, the values ${ \Delta x = \Delta t = 0.005 }$ and ${ l = 100 }$ are kept (except for the last set of simulations, in which ${ l = 200 }$), while ${ t_{\rm{max}} = 150 }$ is sometimes selected to shorten execution times. The other parameters are ${ \sigma_{\rm{on}} = 10 }$, ${ \sigma_{\rm{off}} = 2.5 }$, ${ \alpha = 0.02 }$, and ${ \gamma = 0 }$. Moreover, due to the extent of the computational task, the following investigation focuses on $ \Omega $ in the range ${ \left[ 0.4 , 0.6 \right] }$ and $ A $ in the range ${ \left[ 2 , 2.5 \right] }$ with increments ${ \Delta \Omega = \Delta A = 0.01 }$. Such a restriction is chosen because it fully contains a deterministic breather-only area, along with the edges of the confining ones [see the blue box in Fig.~\ref{fig:2}(a)].

The panels of Fig.~\ref{fig:11} show the results obtained in the presence of the stochastic perturbation ${ \gamma_T (x, t) }$. More specifically, denoting with ${ N_{\rm{b-only}} }$ the number of identified breather-only cases and with $ N $ the total number of realizations, a breather-only generation probability is defined here as ${ P_{\rm{b-only}} = \lim_{N \to \infty} N_{\rm{b-only}} / N }$. Estimations of such a quantity over ${ N = 512 }$ realizations, according to the phase criterion [Eq.~\eqref{eqn:11}], are presented for the noise amplitudes ${ \Gamma = 0.005 }$, ${ 0.05 }$, and ${ 0.5 }$~\cite{Guarcello_2017}. A first remark concerning these plots is that the breather-only regions are still clearly recognizable, since the generation probability quickly rises in their vicinity, see the black/blue points (high-probability cores) in Fig.~\ref{fig:11}. As the value of $ \Gamma $ is increased, a progressively smaller area of high probability is expected to survive because of the larger current fluctuations that tend to break up the breather state, and this is indeed observed. However, a sort of noise-induced widening of these structures is seen as well, because most of the neighboring $ \OA $ combinations, that deterministically are not associated with the formation of breather modes only [i.e., within the dark gray areas in Fig.~\ref{fig:2}(a)], acquire nonzero probability values. Eventually, for rather high noise intensities, ${ \Gamma \sim 1 }$, different kinds of excitations (including breathers, kinks, and antikinks) begin to appear in the medium, in addition to the NST-induced ones, and the breather-only generation probability uniformly falls to zero in the $ \OA $ parameter space.

\begin{figure*}[t!!]
\centering
\includegraphics[width=0.5\textwidth]{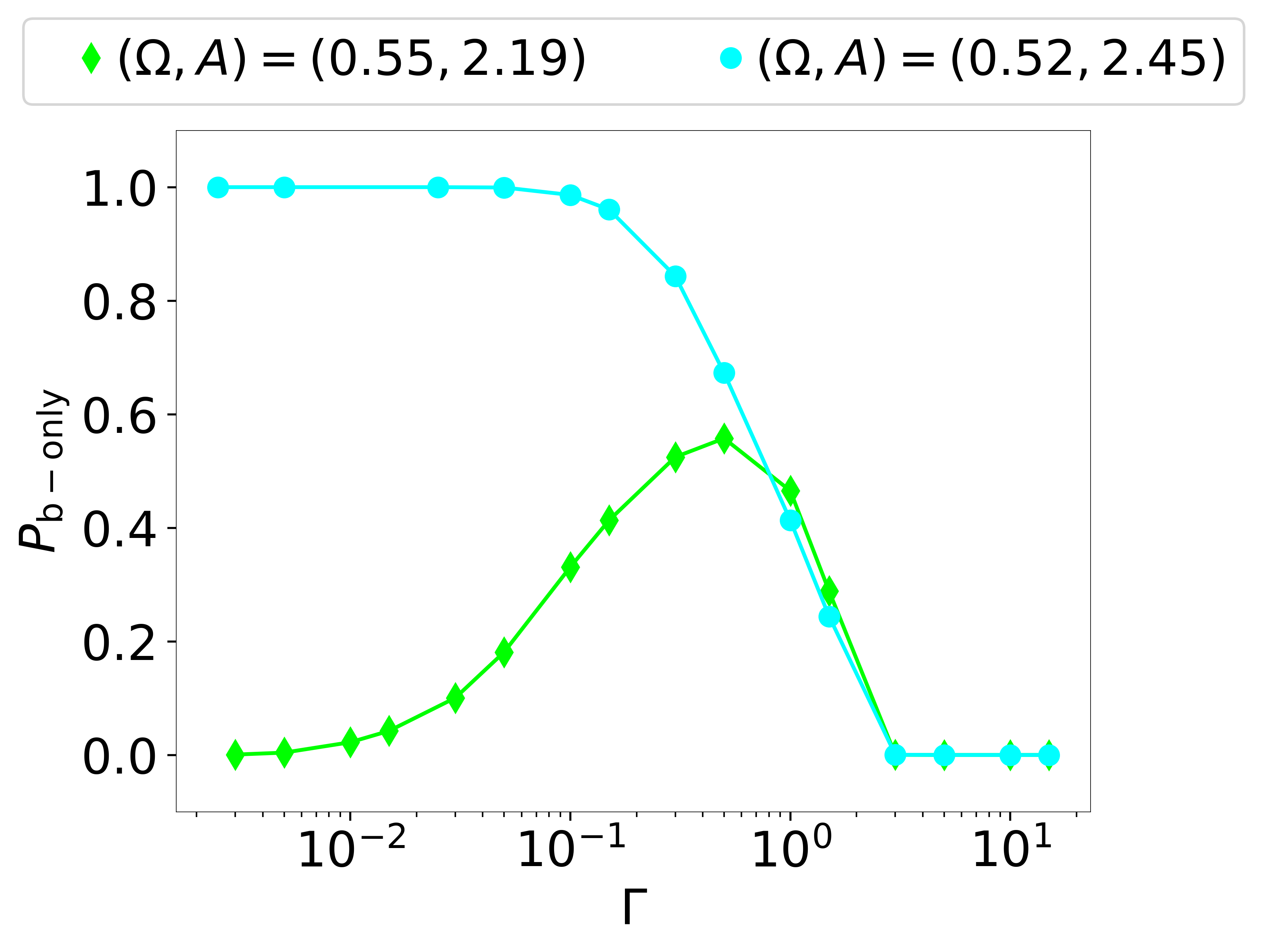}
\caption{Breather-only generation probability estimated over $ N = 10^4 $ realizations, according to the phase criterion [Eq.~\eqref{eqn:11}], as a function of the noise amplitude $ \Gamma $. The green (cyan) color indicates the combination ${ \Omega = 0.55 }$ and ${ A = 2.19 }$ (${ \Omega = 0.52 }$ and ${ A = 2.45 }$). Other parameter values: ${ l = 100 }$, ${ t_{\rm{max}} = 200 }$,  ${ \alpha = 0.02 }$, ${ \gamma = 0 }$, ${ \sigma_{\rm{on}} = 10 }$, and ${ \sigma_{\rm{off}} = 2.5 }$.}
\label{fig:12}
\end{figure*}

To further illustrate the behavior of the probability corresponding to $ \OA $ combinations that do not belong to the breather-only region identified for $ \Gamma = 0 $, the sample point ${ \Omega = 0.55 }$ and ${ A = 2.19 }$ is chosen (see Fig.~\ref{fig:11}), and the breather-only generation probability is calculated for ${ N = 10^4 }$ realizations. In confirmation of the previous considerations, this quantity is a nonmonotonic function of the noise amplitude; as shown in Fig.~\ref{fig:12} (see the green diamonds), the probability is close to zero when ${ \Gamma \to 0 }$, it reaches a peak of approximately ${ 0.6 }$ for an optimal value of $ \Gamma $, and it then goes back to zero when the stochastic influence becomes disruptive, i.e., when the fluctuations are sufficiently strong to both easily break the oscillatory bound state and produce additional modes into the junction.

In a similar fashion, the cyan circles in Fig.~\ref{fig:12} display the breather-only generation probability as a function of the noise amplitude for ${ \Omega = 0.52 }$ and ${ A = 2.45 }$, which is taken as a representative case for the high-probability core, see Fig.~\ref{fig:11}, with ${ N = 10^4 }$. Here, a decreasing steplike profile is observed.

\begin{figure*}[t!!]
\centering
\includegraphics[width=\textwidth]{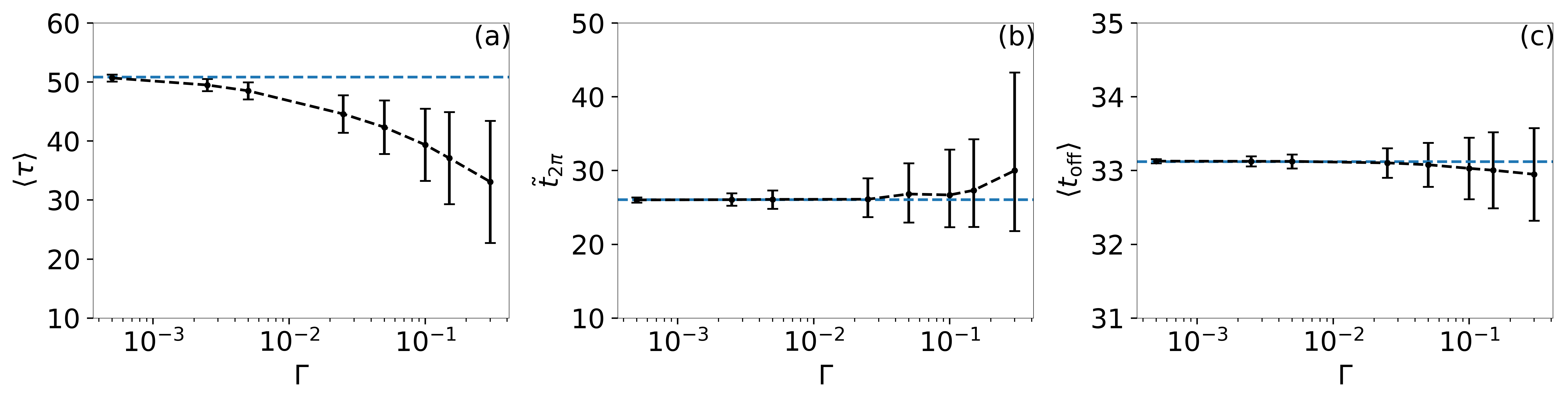}
\caption{Panel~(a): Average breather decay lifetime ${ \left\langle \tau \right\rangle }$ and its standard deviation ${ \sigma_{\tau} }$. Panel~(b): Median of the quantity ${ t_{ 2 \pi } }$ with quartiles. Panel~(c): Average excitation time ${ \left\langle t_{\rm{off}} \right\rangle }$ and its standard deviation ${ \sigma_{t_{\rm{off}}} }$. The quantities in panels~(a),~(b),~and~(c) are represented as a function of the noise amplitude $ \Gamma $, and they are calculated over ${ N_{\rm{b-only}} ( \Gamma ) }$ favorable cases out of ${ N = 10^3 }$ total realizations, see Fig.~\ref{fig:12}. Furthermore, in each plot the dashed blue line indicates the deterministic counterpart, i.e., the corresponding estimation for ${ \Gamma = 0 }$. Parameter values: ${ l = 200 }$, ${ t_{\rm{max}} = 400 }$, ${ \alpha = 0.02 }$, ${ \gamma = 0 }$, ${ \sigma_{\rm{on}} = 10 }$, ${ \sigma_{\rm{off}} = 2.5 }$, ${ \Omega = 0.52 }$, and ${ A = 2.45 }$.}
\label{fig:14}
\end{figure*}

To address the NST-induced breather's dynamics, the average characteristic times ${ \left\langle \tau \right\rangle }$ and ${ \left\langle t_{\rm{off}} \right\rangle }$ are computed in the noisy regime, together with their standard deviations ${ \sigma_{\tau} }$ and ${ \sigma_{t_{\rm{off}}} }$. The quantity ${ t_{2 \pi} }$ is considered as well, but due to the asymmetry of the resulting histograms, the median ${ {\tilde{t}}_{2 \pi} }$ and the quartiles are evaluated in this case. The estimations presented in Fig.~\ref{fig:14} are performed over ${ N_{\rm{b-only}} ( \Gamma ) }$ favorable cases out of $ N $ total realizations, for different values of the noise amplitude (the fact that ${ N_{\rm{b-only}} }$ is $ \Gamma $-dependent is evident from the definition of ${ P_{\rm{b-only}} }$ and from Fig.~\ref{fig:12}). In this framework, it is clearly convenient to work with $ \OA $ combinations belonging to the high-probability core of a breather-only region, hence the pair ${ \Omega = 0.52 }$ and ${ A = 2.45 }$ is maintained. It is also reasonable to first concentrate on the set of noise intensities corresponding to ${ P_{\rm{b-only}} ( \Gamma ) \sim 1 }$. Thus, in the following ${ \Gamma }$ takes values in the range ${ \left[ 0.0005 , 0.3 \right] }$ (see Fig.~\ref{fig:12}).

Two main qualitative facts emerging from the simulations should be highlighted. First, as the value of the noise intensity $ \Gamma $ is increased, the time series ${ \varphi_{\rm{max}} (t) }$ [defined in Eq.~\eqref{eqn:14}] is progressively seen to differ more from one realization to another: rapid extinctions can occur, but in other cases the breather seems quite robust against the current fluctuations. Secondly, for ${ \Gamma > 0 }$ the breather's decaying amplitude is trackable down to the fluctuation level, but obviously not below it. Therefore, the fitting curve of Eq.~\eqref{eqn:21} is replaced by a slightly different one, i.e.,
\begin{equation}
\label{eqn:22}
h (t) = \begin{cases}
	B - C & t_{\rm{off}} \leq t < t_{\rm{off}} + t_{ 2 \pi } \\
	\left( B - C \right) {\rm{e}}^{ - \frac{ t - \left( t_{\rm{off}} + t_{ 2 \pi } \right) } {\tau} } + C & t \geq t_{\rm{off}} + t_{ 2 \pi } ,
 \end{cases}
\end{equation}
with $ \tau $ now measuring the damped breather amplitude's persistence time above the (approximate) fluctuation level ${ C > 0 }$, the latter being a fitting parameter as well. Here, it is assumed that an exponential-like trend is still able to follow the breather's dissipative relaxation process, which appears to be fairly true for the set of noise intensities under consideration. Also, note that $ \tau $ matches the deterministic value, computed in Sec.~\ref{Sec3a}, for ${ C = 0 }$, and that the definitions of ${ t_{ 2 \pi } }$ and ${ t_{\rm{off}} }$ remain the same as before.

In order for a procedure based on Eq.~\eqref{eqn:22} to work, $ C $'s estimation should be carried out under appropriate conditions. In particular, the simulation time ${ t_{\rm{max}} }$ should be long enough to let ${ \varphi_{\rm{max}} (t) }$'s local maxima unequivocally reach the previously defined fluctuation level without any reflection effects at ${ x = l }$. This motivates the choices of ${ l = 200 }$, ${ t_{\rm{max}} = 400 }$, and ${ N = 10^3 }$ for the last set of simulations.

The obtained average breather decay lifetime ${ \left\langle \tau \right\rangle }$ and its standard deviation ${ \sigma_{\tau} }$ are displayed in Fig.~\ref{fig:14}(a) for different noise amplitudes. A significant degree of robustness can be deduced from the plot, since there exists a range of $ \Gamma $ values where ${ \left\langle \tau \right\rangle }$ remains close to its deterministic counterpart [indicated by the dashed blue line in Fig.~\ref{fig:14}(a)]. As the influence of the stochastic perturbation increases, the breather's dissipative relaxation process varies more between different realizations; in most cases it occurs faster, also due to the fact that the fluctuation level ${ C ( \Gamma ) }$ monotonically grows, hence the average's decrease. However, the noise is sometimes able to sustain the oscillatory excitation quite effectively, and it can lead to values ${ \tau \gtrsim 1 / \alpha }$ (not explicitly shown here), which is very interesting.

Lastly, Figs.~\ref{fig:14}(b)~and~\ref{fig:14}(c) present, respectively, the results concerning the quantities ${ {\tilde{t}}_{2 \pi} }$ and ${ \left\langle t_{\rm{off}} \right\rangle }$. As already mentioned, the randomly fluctuating term will eventually lead the soliton towards the kink-antikink state, therefore the monotonically increasing profile of Fig.~\ref{fig:14}(b) is a reasonable outcome. As for the average excitation time, its closeness to the deterministic value suggests that, once the typical duration of a pulse yielding a single breather is found for ${ \Gamma = 0 }$, it still constitutes a reliable choice for a wide range of noise intensities.

\section{Conclusions}
\label{Sec4}

The emergence of travelling breather modes in a driven LJJ is numerically studied by looking at their localization in the $ \OA $ parameter space, where $ \Omega $ and $ A $ indicate, respectively, the frequency and the amplitude of the modulated magnetic pulse which forces the junction at one end. By means of the two detection criteria introduced here, refined bifurcation diagrams are produced under various conditions. This allows for the identification of the $ \OA $ combinations for which the NST generation process results in the formation of breather excitations only. Furthermore, quantities such as the medium's energy and the breather radiative decay lifetime are taken into consideration to characterize the physics of these nonlinear excitations.

In particular, this paper documents the existence of significant breather-only regions in the $ \OA $ parameter space, whose extension and structure depends on the typical duration of the switching-on/off regimes of the driving pulse, the damping coefficient, and the current biasing the junction. For example, additional kinks (and/or antikinks) are progressively seen to enter the medium as the typical time scale of the switching-off regime of the forcing signal is incremented. Also, for higher values of the damping parameter, gradually larger breather-only areas are obtained. However, the resulting excitations are shorter-lived, in accordance with the perturbative prediction on the breather radiative decay lifetime. Lastly, the breather-only regions tend to disappear as the applied bias current increases, since the latter term tends to tear the kink-antikink couple apart.

Then, a white Gaussian noise source is included to check whether the breather-only zones identified in the deterministic case are robust enough to survive in a randomly fluctuating environment. By defining a probability of breather-only generation, it is seen that these areas essentially maintain their identity in the presence of the stochastic perturbation. In fact, for a vast set of noise intensities, a core of high probability can be clearly recognized in correspondence to the $ \OA $ values that deterministically lead to the emergence of breather modes only. Moreover, a sort of positive noise-induced effect is found, since nonzero probability values appear outside of the deterministic breather-only regions. Interestingly, such a fact is seen to occur in correspondence with a nonmonotonic behavior of the breather-only generation probability as a function of the noise amplitude: the probability is close to zero when the dynamics is closer to the deterministic case, it reaches its peak for an optimal value of the noise intensity, and it eventually goes back to zero when the stochastic influence becomes disruptive. In addition, the breather seems to stand up the stochastic background quite well, since there exists a range of noise amplitudes values where its average persistence time above the fluctuation level remains close to the inverse of the damping coefficient. The latter result represents a first step towards the characterization of the stochastic dynamics of breathers, which is still a largely unexplored topic.

\section*{Acknowledgments}
\label{acknowledgments}

This work was supported by Project R4D08-P3PGDAR4 \textunderscore MARGINE, Università degli Studi di Palermo; the Government of the Russian Federation through Agreement No. 074-02-2018-330 (2); and the Italian Ministry of University and Research (MUR).


\end{document}